\documentclass[a4paper, oneside, twocolumn,
  notitlepage, 10pt, final]{extarticle_ecoc}
\usepackage{ecoc}

\begin{document}
\selectlanguage{english}

%--- Title ---%

\title{
Modeling the Impact of Fiber Latency on
  Compute-Communication Overlap
  in Geo-Distributed Multi-Datacenter AI Training}

%--- Authors ---%

\author{
    Ioannis Papavasileiou\textsuperscript{(1)},
    Sairam Prabhakar\textsuperscript{(1)},
    Indu Kant Deo\textsuperscript{(1)}, 
    Sergejs Makovejs\textsuperscript{(1)}
}

\maketitle

%--- Author Description ---%

\begin{strip}
    \begin{author_descr}

        \textsuperscript{(1)} Corning Inc., Corning, New York, USA \textcolor{blue}{\uline{papavasii@corning.com}}

    \end{author_descr}
\end{strip}

%--- Footnote ---%
\renewcommand\footnotemark{}
\renewcommand\footnoterule{}

%--- Abstract ---%

\begin{strip}
    \begin{ecoc_abstract}
        We use discrete-event simulation to quantify the impact of fiber latency on 
the efficacy of geo-distributed AI model training with data parallelism.
We conclude that 
the optimum distances between two AI clusters is 10-100km, over which hollow-core fiber
enables 25\% higher compute-communication overlap.
        \textcopyright~2026 The Author(s)
    \end{ecoc_abstract}
\end{strip}

%--- Introduction ---%

\section{Introduction}

The current generation of Large Language Models (LLMs)
feature on the order of $10^{12}$ learnable parameters
trained on $10^{13}$ tokens of text data
\cite{ref:gpt3, ref:gpt4}.
Training models of this scale is not feasible on a
single Graphical Processing Unit (GPU) due to memory
limitations and prohibitively long training times.
As a result, distributed computation across multiple
GPUs has become essential in artificial intelligence
(AI) model engineering.
Moreover, power delivery constraints and physical
infrastructure limitations increasingly prevent the
concentration of sufficiently large GPU clusters
within a single datacenter \cite{ref:dc_power},
while building facilities in remote locations with
abundant and affordable power further motivates
geographic distribution. Together, these factors are
driving the adoption of multi-datacenter training
architectures where computation is distributed across
geographically separated sites.
In such distributed systems, efficient communication
between GPUs is critical for achieving high training
throughput.

Data parallelism is one strategy for
distributed training, where the model is replicated
across multiple GPUs and each GPU processes a different
subset of the training data. After computing local
gradients, GPUs must synchronize through collective
communication operations such as all-reduce to average
gradients before updating model parameters
\cite{ref:blueconnect}. Modern deep learning frameworks
employ compute-communication overlap techniques, where
gradient synchronization is performed concurrently with
 computation to hide communication latency
\cite{ref:pytorch_distributed, ref:co2overlap}.
Discrete-event simulation (DES) methodology enables the analysis of transient 
compute and communication events during LLM training across various network configurations.

However, as network interface card (NIC) bandwidths
continue to increase---from 400\,Gbps to 800\,Gbps
and beyond 1.6\,Tbps---the serialization time for data
transmission decreases, causing fiber propagation
latency to become an increasingly dominant factor in
total communication delay. For geo-distributed training
across datacenters (DC) separated by tens to hundreds of
kilometers, this latency can significantly degrade
compute-communication overlap and reduce effective GPU
utilization. Despite the growing interest in
scale-across multi-DC training architectures,
the impact of fiber latency on training efficiency has
not been systematically characterized.

In this paper, we use DES to simulate
data-parallel training of GPT-3 models (13B and 175B
parameters) across dual-DC GPU clusters of up
to 8192 units, comparing hollow core fiber (HCF) and
standard single-mode fiber (SMF) over distances from
0.3\,km to 1000\,km. To the best of our knowledge,
this is the first systematic study that:
(1)~quantifies how inter-DC fiber propagation
latency degrades compute-communication overlap as a
function of distance and GPU generation; and
(2)~demonstrates that HCF enables up to 25\% higher
compute-communication overlap than SMF at
inter-DC distances of 10--100\,km.

%--- Methods ---%

\section{Methods}

\textbf{Communication Time Model.}
For point-to-point communication between GPUs, the
total time to transmit a message can be decomposed
into serialization time (bandwidth-dependent) and
propagation time (latency-dependent):
\begin{equation}\label{eq:comm_time}
    T_{\mathrm{comm}}
      = T_{\mathrm{serialization}}
      + T_{\mathrm{propagation}}
      = \frac{M}{B} + \frac{D}{v}
\end{equation}
where $M$ is the message size, $B$ is the NIC
bandwidth, $D$ is the distance between GPUs, and $v$
is the signal propagation speed in the fiber. For SMF,
$v \approx 2 \times 10^{8}$\,m/s, while HCF achieves
$v \approx 3 \times 10^{8}$\,m/s (close to the speed
of light in vacuum), resulting in $\sim33\%$ lower propagation latency.
We assume negligible switch latency  and dedicated uncongested LLM training network.

As NIC bandwidths increase, the serialization term
decreases, causing the propagation term to dominate
for messages transmitted over longer distances.
Gradient bucket sizes in distributed training
typically range from 1--100\,MB, with most collective communication frameworks providing bucket
sizes in this range \cite{ref:pytorch_distributed}. Pure data-parallel
workloads operate at the upper end, while hybrid
parallelism strategies that combine model and data parallelism
\cite{ref:megatron, ref:megatron_sc21} reduce the
per-rank gradient volume and push effective message
sizes toward the lower end. This can cause 
the communication to enter a latency-dominated regime
at longer distances.

\textbf{Compute-Communication Overlap.}
Distributed training efficiency depends on
the ability to overlap communication with computation.
During backpropagation, gradients are computed
layer-by-layer and can be communicated incrementally
or combined through bucketing. While one chunk
undergoes all-reduce synchronization, the GPU
continues computing gradients for subsequent layers. The degree of overlap is
quantified as:
\begin{equation}\label{eq:overlap}
    \eta_{\mathrm{overlap}}
      = \frac{T_{\mathrm{compute}}}{T_{\mathrm{total}}}
\end{equation}
where $T_{\mathrm{compute}}$ is the time spent on
computation and $T_{\mathrm{total}}$ is the total
training iteration time. When
$\eta_{\mathrm{overlap}} = 1$, all communication is
hidden behind compute \cite{ref:co2overlap}; when
$\eta_{\mathrm{overlap}} < 1$, GPUs experience idle
time waiting for communication to complete.

\textbf{ASTRA-sim.}
For our simulations we use ASTRA-sim, an open-source DES for distributed machine learning systems
\cite{ref:astrasim1, ref:astrasim2}.
It models the interplay between training workload
 (model architecture, compute
requirements), system configuration (collective
communication algorithms), network properties
(topology, bandwidth, latency),
and schedules events to enforce compute-communication overlap. 

\begin{figure}[t!]
    \centering
    \includegraphics[width=0.8\columnwidth]{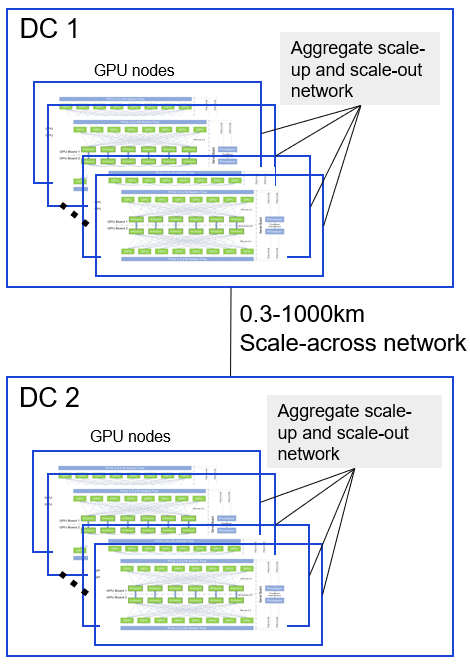}
    \caption{Multi-DC simulation architecture
      with lumped DC representation. Each DC contains
      up to 4096 GPUs with high-bandwidth intra-DC
      connectivity; inter-DC communication occurs over
      SMF or HCF links at variable distances.}
    \label{fig:architecture}
\end{figure}

\textbf{Lumped DC Representation.}
To focus on inter-DC communication effects
and simplify the intra-DC complexities,
we adopt a lumped DC abstraction where
each DC is represented as a single high-capacity GPU
node with symmetric internal connectivity. This simplification avoids modeling detailed
leaf-spine network topologies, which means our results isolate the impact
of inter-DC propagation latency. Figure~\ref{fig:architecture} illustrates
the multi-DC architecture with two lumped DC nodes
connected via long-haul fiber.

\textbf{Simulation Parameters.}
Table~\ref{tab:sim_params} summarizes the simulation
parameter space. We simulate two DCs with
data-parallel training of GPT-3 models (13B and 175B
parameters) \cite{ref:gpt3}.
We estimate compute times from theoretical peak GPU
performance: 312 TFLOPS for A100 and 989 TFLOPS for
H100 (tensor cores). Intra-DC communication uses a
switch topology with 6 parallel links, representing 
effective attributes for an aggregated representation 
of NVLink (scale-up) and InfiniBand/Ethernet (scale-out) fabric within the DC. 
Inter-DC bandwidth is set to 100\,GB/s or 200\,GB/s,
 with latency determined by distance and fiber type.

\begin{table}[h!]
    \centering
    \caption{Simulation parameter space}
    \label{tab:sim_params}
    \begin{tabular}{|l|l|}
        \hline
        \textbf{Parameter} & \textbf{Values} \\
        \hline
        Model & GPT-3 13B, 175B \\
        \hline
        Total GPUs & 256, 2048, 8192 \\
        \hline
        GPU type & NVIDIA A100, H100 \\
        \hline
        Fiber type & HCF, SMF \\
        \hline
        Inter-DC distance & 0.3\,km -- 1000\,km \\
        \hline
        Inter-DC bandwidth & 100\,GB/s, 200\,GB/s$^{*}$ \\
        \hline
        Intra-DC bandwidth & 600\,GB/s (6$\times$100\,GB/s) \\
        \hline
        Intra-DC latency & 1\,$\mu$s \\
        \hline
        \multicolumn{2}{|l|}{\footnotesize
          $^{*}$200\,GB/s tested for 256 GPU
          configuration only} \\
        \hline
    \end{tabular}
\end{table}

%--- Results ---%

\section{Results}

\begin{figure}[t!]
    \centering
    \includegraphics[width=\columnwidth]{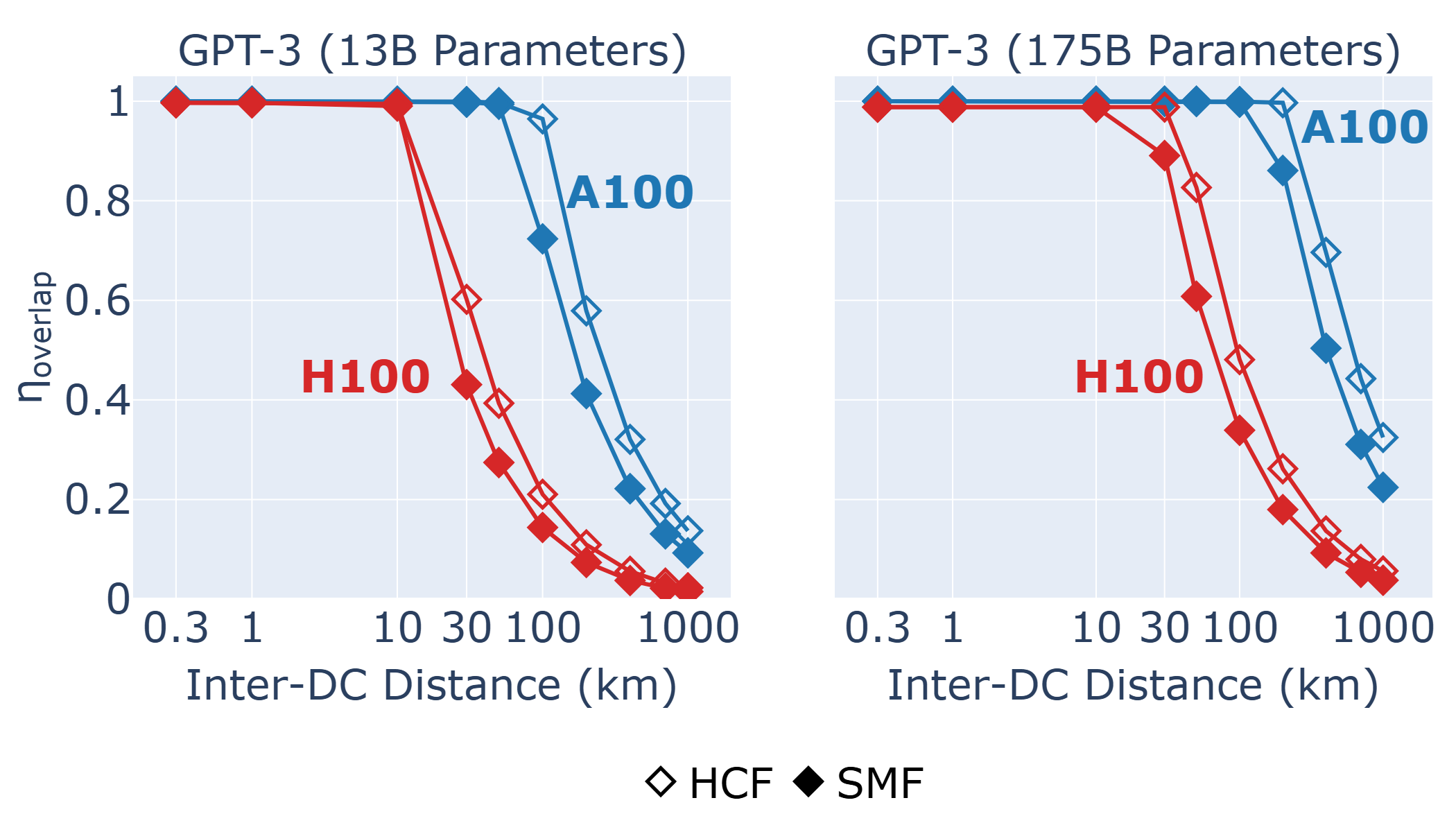}
    \caption{Compute-communication overlap
      ($\eta_{\mathrm{overlap}}$) vs.\ inter-DC distance
      for 8192 GPUs. Left: GPT-3 13B; right: GPT-3 175B.
      hollow markers: HCF; filled markers: SMF. Blue: A100;
      red: H100. HCF consistently achieves higher overlap
      than SMF across all configurations.}
    \label{fig:overlap_results}
\end{figure}

Figure~\ref{fig:overlap_results} presents the
compute-communication overlap
($\eta_{\mathrm{overlap}}$) as a function of
inter-DC distance for the 8192 GPU
configuration. 

Several key trends emerge from the simulations. First,
compute-communication overlap decreases monotonically
with increasing inter-DC distance for all
configurations, as longer propagation delays reduce
the ability to mask communication behind computation.
At distances below 10\,km, near-complete overlap
($\eta_{\mathrm{overlap}} \approx 1$) is achieved
regardless of fiber type. As the distance increases, overlap drops significantly for both fiber
types, though HCF maintains a consistent advantage.

\begin{figure}[t!]
    \centering
    \includegraphics[width=.98\columnwidth]{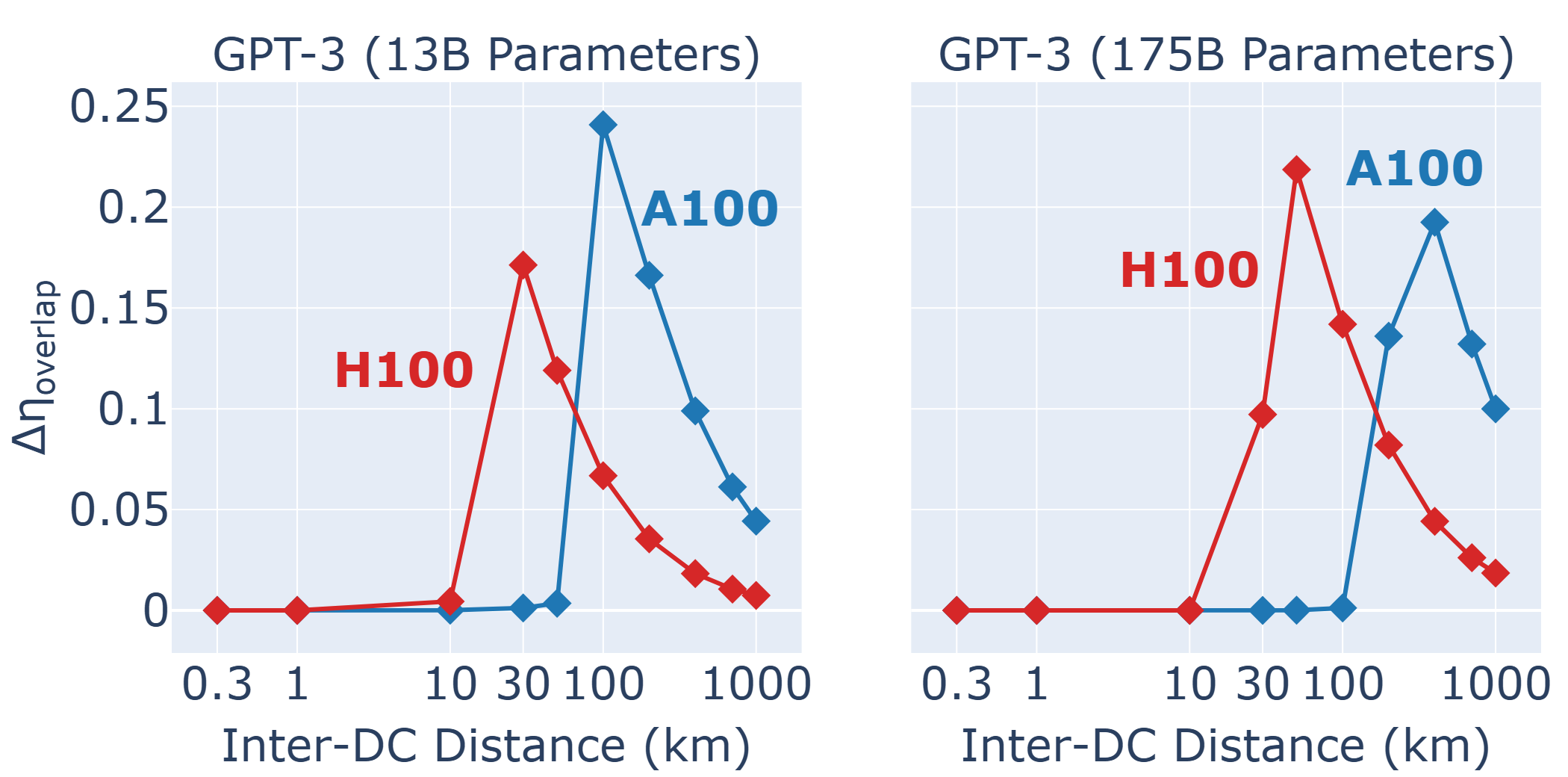}
    \caption{Absolute improvement in
      compute-communication overlap for HCF over SMF
      ($\Delta\eta = \eta_{\mathrm{HCF}}
        - \eta_{\mathrm{SMF}}$)
      vs.\ inter-DC distance for 8192 GPUs.
      Left: GPT-3 13B; right: GPT-3 175B.
      Blue: A100; red: H100. The benefit peaks at
      intermediate distances where the system
      transitions from bandwidth-limited to
      latency-limited.}
    \label{fig:hcf_benefit}
\end{figure}

Second, HCF outperforms SMF across all tested
configurations. Figure~\ref{fig:hcf_benefit} shows
the absolute improvement in overlap
($\Delta\eta = \eta_{\mathrm{HCF}}
  - \eta_{\mathrm{SMF}}$)
with the benefit peaking at
intermediate distances where the system transitions
from bandwidth-limited to latency-limited operation.
In the latency-dominated regime, HCF can
reach up to 50\% greater inter-DC separation than
SMF for the same propagation delay, extending the
feasible radius for multi-DC deployments.

\begin{figure}[t!]
    \centering
    \includegraphics[width=\columnwidth]{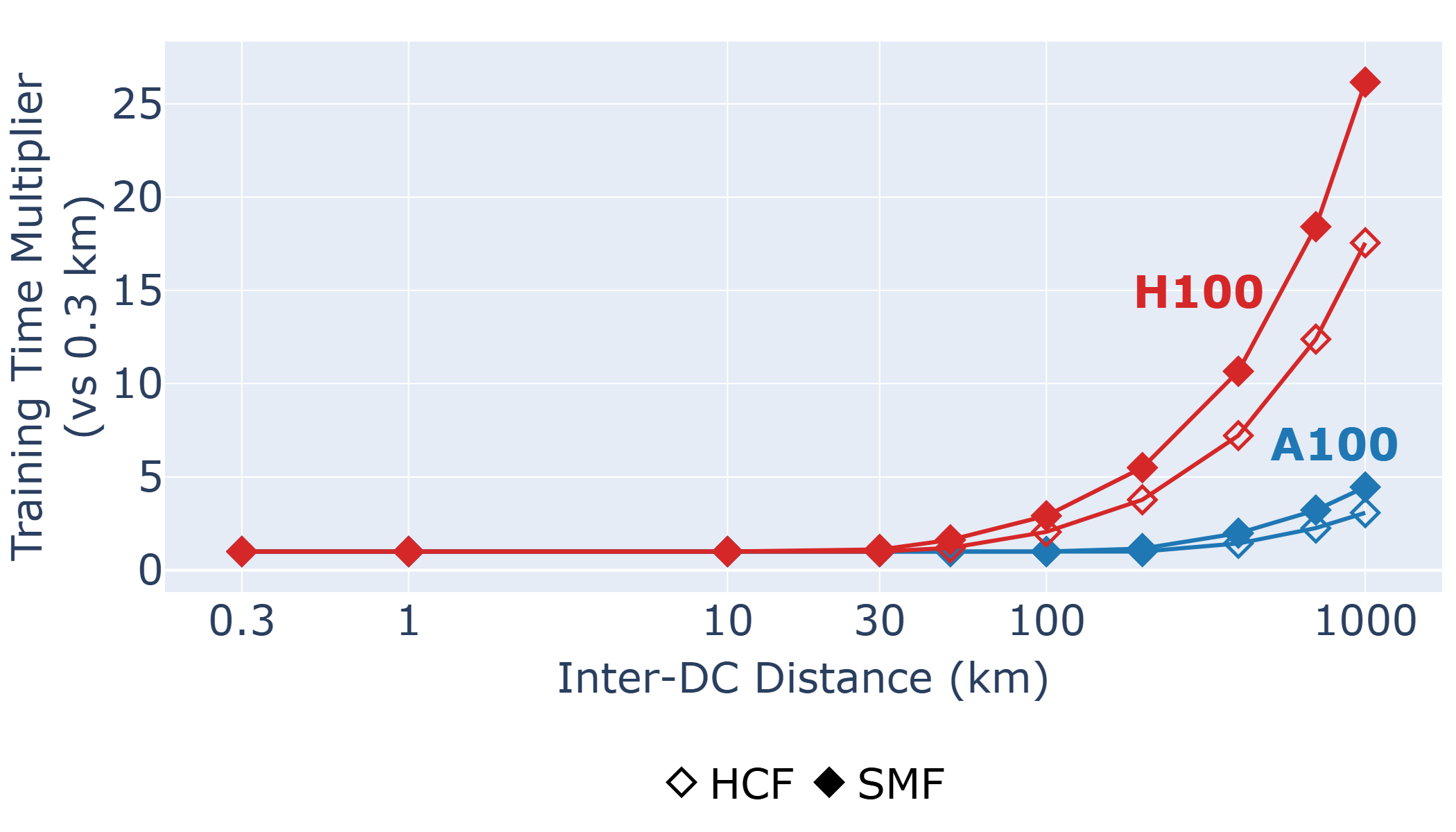}
    \caption{Training time multiplier relative to a 0.3\,km baseline
      for GPT-3 175B on 8192 GPUs.
      Blue: A100; red: H100.
      Filled markers: SMF; hollow markers: HCF.
      H100 is significantly more sensitive to inter-DC distance
      due to its higher compute throughput,
      reaching ${\sim}26\times$ at 1000\,km on SMF
      versus ${\sim}4\times$ for A100.
      HCF reduces the H100 penalty from
      $26\times$ to ${\sim}17\times$.}
    \label{fig:training_time_multiplier}
\end{figure}

However, at very long distances both fiber types yield low
absolute overlap, limiting practical utility (figure~\ref{fig:overlap_results}). 
This low overlap is affecting the total training time, 
as captured in figure~\ref{fig:training_time_multiplier}. 
Training time increases with distance, 
with H100 being impacted more at longer distances,
due to the less overlap when compared to A100.
This further validates that 
all-reduce operations with communication
message size and frequency to maximize 
compute-communincation overlap at longer distances 
become very costly.

Third, larger models exhibit higher baseline overlap
than smaller models. The GPT-3 175B model, with its
greater computational requirements per layer, provides
more time for communication to complete in the
background compared to the 13B variant. This suggests
that the largest frontier models may be more amenable
to geo-distributed training than smaller models.

Fourth, faster GPUs reduce overlap by completing
computation more quickly, thereby exposing
communication latency. The H100, which is
approximately 3$\times$ faster than the A100 in FP16
tensor operations, shows lower overlap values across
all distances. Note that lower overlap does not imply slower
training---H100 clusters still complete iterations
faster in absolute time---but it does indicate lower
hardware efficiency for these GPT-3 transformer models. As GPU compute capabilities
continue to advance, communication latency will
become an increasingly critical bottleneck.

Simulations with 256 and 2048 GPUs exhibit the same
trends in overlap efficiency and are omitted for
brevity. To isolate the role of bandwidth versus
latency, we also doubled the inter-DC link rate from
800\,Gbps to 1.6\,Tbps for the 256-GPU
configuration. When bandwidth is doubled, 
a maximum of 0.66\%
improvement is observed in $\eta_{\mathrm{overlap}}$,
confirming that propagation latency---not
bandwidth---is the dominant bottleneck at
inter-DC distances for communication message sizes aimed at
maximizing compute-communication overlap.

%--- Conclusions ---%

\section{Conclusions}

ASTRA-sim simulations of data-parallel GPT-3 training
show that compute-communication overlap degrades
significantly with inter-DC distance. HCF
consistently outperforms SMF when performing synchronous per-layer collective communications
and it enables up to 50\% greater separation and 25\% higher compute-communication overlap.
Larger models and
slower GPUs are more amenable to geo-distribution
due to higher compute-to-communication ratios.
This benefit is more pronounced between 10-100km distances 
where the overlap benefits are high but the training time 
degredation due to latency-sensitive communication is low.
Although not studied in this work, we expect that the use of hybrid (data and model) 
parallelization strategies \cite{ref:megatron, ref:megatron_sc21} 
will reduce 
per-message sizes and enter the latency-dominated
regime at shorter distances.
This, will expand the range of distances where the benefit of low-latency fiber is prevalent. 
Future work will extend to hybrid parallelism,
mixture-of-experts architectures, distributed
inference workloads, 
bucketing of gradient updates,
and multi-tier network hierarchies.

\clearpage

%--- Bibliography ---%

\printbibliography

@article{ref:gpt4,
  author  = "OpenAI",
  title   = "{GPT-4} Technical Report",
  journal = "arXiv preprint arXiv:2303.08774",
  year    = "2023",
  doi     = "10.48550/arXiv.2303.08774"
}

@inproceedings{ref:astrasim1,
  author    = "Saeed Rashidi and Srinivas Sridharan and Sudarshan Srinivasan and Tushar Krishna",
  title     = "{ASTRA-sim}: Enabling {SW/HW} Co-Design Exploration for Distributed {DL} Training Platforms",
  booktitle = "Proceedings of the IEEE International Symposium on Performance Analysis of Systems and Software (ISPASS)",
  pages     = "81--92",
  year      = "2020",
  doi       = "10.1109/ISPASS48437.2020.00018"
}

@inproceedings{ref:astrasim2,
  author    = "William Won and Taekyung Heo and Saeed Rashidi and Srinivas Sridharan and Sudarshan Srinivasan and Tushar Krishna",
  title     = "{ASTRA-sim2.0}: Modeling Hierarchical Networks and Disaggregated Systems for Large-model Training at Scale",
  booktitle = "Proceedings of the IEEE International Symposium on Performance Analysis of Systems and Software (ISPASS)",
  address   = "Raleigh, NC, USA",
  pages     = "283--294",
  year      = "2023",
  doi       = "10.1109/ISPASS57527.2023.00035"
}

@article{ref:blueconnect,
  author  = "Minsoo Cho and Ulrich Finkler and Marcelo Serrano and David Kung and Hillery Hunter",
  title   = "{BlueConnect}: Decomposing All-Reduce for Deep Learning on Heterogeneous Network Hierarchy",
  journal = "IBM Journal of Research and Development",
  volume  = "63",
  number  = "6",
  pages   = "1:1--1:11",
  year    = "2019",
  doi     = "10.1147/JRD.2019.2947013"
}

@article{ref:co2overlap,
  author  = "Weigao Sun and Zhen Qin and Weixuan Sun and Shidi Li and Dong Li and Xuyang Shen and Yu Qiao and Yiran Zhong",
  title   = "{CO2}: Efficient Distributed Training with Full Communication-Computation Overlap",
  journal = "arXiv preprint arXiv:2401.16265",
  year    = "2024",
  doi     = "10.48550/arXiv.2401.16265"
}

@article{ref:pytorch_distributed,
  author  = "Shen Li and Yanli Zhao and Rohan Varma and Omkar Salpekar and Pieter Noordhuis and Teng Li and Adam Paszke and Jeff Smith and Brian Vaughan and Pritam Damania and Soumith Chintala",
  title   = "{PyTorch} Distributed: Experiences on Accelerating Data Parallel Training",
  journal = "arXiv preprint arXiv:2006.15704",
  year    = "2020",
  doi     = "10.48550/arXiv.2006.15704"
}

@article{ref:gpt3,
  author  = "Tom Brown and Benjamin Mann and Nick Ryder and Melanie Subbiah and Jared Kaplan and Prafulla Dhariwal and Arvind Neelakantan and Pranav Shyam and Girish Sastry and Amanda Askell and others",
  title   = "Language Models are Few-Shot Learners",
  journal = "Advances in Neural Information Processing Systems",
  volume  = "33",
  pages   = "1877--1901",
  year    = "2020"
}

@article{ref:megatron,
  author  = "Mohammad Shoeybi and Mostofa Patwary and Raul Puri and Patrick LeGresley and Jared Casper and Bryan Catanzaro",
  title   = "{Megatron-LM}: Training Multi-Billion Parameter Language Models Using Model Parallelism",
  journal = "arXiv preprint arXiv:1909.08053",
  year    = "2020",
  doi     = "10.48550/arXiv.1909.08053"
}

@inproceedings{ref:megatron_sc21,
  author    = "Deepak Narayanan and Mohammad Shoeybi and Jared Casper and Patrick LeGresley and Mostofa Patwary and Vijay Anand Korthikanti and Dmitri Vainbrand and Prethvi Kashinkunti and Julie Bernauer and Bryan Catanzaro and Amar Phanishayee and Matei Zaharia",
  title     = "Efficient Large-Scale Language Model Training on {GPU} Clusters Using {Megatron-LM}",
  booktitle = "Proc. Int. Conf. High Performance Computing, Networking, Storage and Analysis (SC)",
  year      = "2021",
  doi       = "10.1145/3458817.3476209"
}

@techreport{ref:dc_power,
  author      = "{International Energy Agency}",
  title       = "Data Centres and Data Transmission Networks",
  institution = "IEA",
  address     = "Paris",
  year        = "2024",
  url         = "https://www.iea.org/energy-system/buildings/data-centres-and-data-transmission-networks"
}

%--- End of Document ---%
\end{document}